\documentclass[submission,copyright,creativecommons]{eptcs}
\providecommand{\event}{Non-Classical Logics 2024 (NCL'24)}

\usepackage{proof}

\newcommand{\PROOF}{\mbox{\bf Proof.}\ \ }
\newcommand{\QED}{\hfill \rule{1.3ex}{0.6em}}
\newcommand{\SEQ}[2]{\mbox{$ #1 \Rightarrow #2 $}}
\newcommand{\I}{\mbox{$\to$}} 
\newcommand{\LAND}{\mbox{$\land$}} 
\newcommand{\LOR}{\mbox{$\lor$}} 
 
\newcommand{\NEG}{\mbox{$\sim$}}
\newcommand{\al}{\mbox{\it $\alpha$}}
\newcommand{\be}{\mbox{\it $\beta$}}
\newcommand{\ga}{\mbox{\it $\gamma$}}
\newcommand{\de}{\mbox{\it $\delta$}}

\newcommand{\GA}{\mbox{\it $\Gamma$}}
\newcommand{\DE}{\mbox{\it $\Delta$}}
\newcommand{\SI}{\mbox{\it $\Sigma$}}

\newcommand{\PH}{\mbox{\it $\Phi$}}

\newtheorem{prop}{Proposition}[section] 
\newtheorem{thm}[prop]{Theorem} 
\newtheorem{lm}[prop]{Lemma} 
 
\newtheorem{df}[prop]{Definition} 
 
\newtheorem{rmk}[prop]{Remark}

\usepackage{iftex}

\ifpdf
  \usepackage{underscore}         % Only needed if you use pdflatex.
  \usepackage[T1]{fontenc}        % Recommended with pdflatex
\else
  \usepackage{breakurl}           % Not needed if you use pdflatex only.
\fi

\title{Unified Gentzen Approach to Connexive Logics over Wansing's C}
\author{Norihiro Kamide
\institute{Nagoya City University, Aichi, Japan}
\email{drnkamide08@kpd.biglobe.ne.jp}
}
\def\titlerunning{Unified Gentzen Approach to Connexive Logics over Wansing's C}
\def\authorrunning{Norihiro Kamide}
\begin{document}
\maketitle

\begin{abstract}
Gentzen-style sequent calculi and Gentzen-style natural deduction systems are introduced for a family (C-family) of connexive logics over Wansing's basic connexive logic C. The C-family is derived from C by incorporating the Peirce law, the law of excluded middle, and the generalized law of excluded middle. Theorems establishing equivalence between the proposed sequent calculi and natural deduction systems are demonstrated. Cut-elimination and normalization theorems are established for the proposed sequent calculi and natural deduction systems, respectively.
\end{abstract}

%%%%%%%%%%%%%%%%%%%%%%%%%%%%%%%%%%%%%%%%%%%%%%
\section{Introduction}

\begin{sloppypar}
Connexive logics are recognized as philosophically plausible paraconsistent logics \cite{ANGELL1962,MCCALL1966,WANSINGAIML2005,WANSING-CN}. A distinguishing feature of connexive logics is their validation of the so-called Boethius' theses: 
$(\al\I\be) \I \NEG(\al \I \NEG\be)$ and 
$(\al \I \NEG \be) \I \NEG(\al\I\be)$. 
On one hand, the roots of connexive logics can be traced back to Aristotle and Boethius. On the other hand, modern perspectives on connexive logics were established by Angell \cite{ANGELL1962} and McCall \cite{MCCALL1966}. 
\end{sloppypar}

\begin{sloppypar}
A basic constructive connexive logic referred to as C, considered a variant of Nelson's paraconsistent logic N4 \cite{AN84,NELSON,KW-BOOK-2015}, was introduced by Wansing in \cite{WANSINGAIML2005}. Additionally, C was extended by Wansing in \cite{WANSINGAIML2005} to introduce a constructive connexive modal logic, serving as a constructive connexive analogue of the smallest normal modal logic K. 
For further details on connexive logics, refer to, for example, \cite{ANGELL1962,MCCALL1966,WANSINGAIML2005,C-NDJFL-2008,KWBOOK2011,KWIFCOLOG2016,WANSING-CN,OW-AiML-2020,NW-JPL-2023} and the references therein.
\end{sloppypar}

In this study, a unified Gentzen-style framework is employed to investigate several connexive logics over Wansing's C. 
%%%%%% 2024/7/12 %%%%%%%%%%%%%%
The term ``unified Gentzen-style framework'' means that we can handle Gentzen-style sequent calculus and Gentzen-style natural deduction system uniformly, with an equivalence between them. 
%%%%%%%%%%%%%%%%%%%%%%%%%%%%%%%
The logics under consideration include Omori and Wansing's connexive logic C3 \cite{OW-AiML-2020}, material connexive logic MC \cite{WANSING-CN}, and Cantwell's connexive logic CN \cite{C-NDJFL-2008}. C3 is obtained from C by adding the law of excluded middle $\neg\al\LOR\al$, MC is obtained from C by adding the Peirce law $((\al\I\be)\I\al)\I\al$, and CN is obtained from C3 by adding the Peirce law.

%%%%%%% 2024/7/14 %%%%%%%%%%%%%%%%
On one hand, Gentzen-style or G3-style sequent calculi for C, C3, CN and some intermediate logics between C and C3 have been introduced and investigated \cite{WANSINGAIML2005,OW-AiML-2020,ERS-2021,Niki-FLAP-2024}, along with a Gentzen-style natural deduction system for the implicational fragment of C \cite{KAMIDE-JANCL-2005}. 
%%%
On the other hand, a unified Gentzen-style framework for C, C3, MC, and CN has not been established. Therefore, we construct such a framework in this study. This framework enables an integrated proof-theoretical treatment of these logics and establishes a natural correspondence between sequent calculi and natural deduction systems for them.
%%%%%%%%%%%%%%%%%%%%%%%%%%%%%%%%%%%

We now discuss some related works on sequent calculi for connexive logics. The cut-elimination theorem for a Gentzen-style sequent calculus, referred to as sC, was proved by Wansing in \cite{WANSINGAIML2005}, although the name sC was not used by him. The cut-elimination theorems for G3-style sequent calculi, namely G3C and G3C3at for C and C3, respectively, were established by Omori and Wansing in \cite{OW-AiML-2020}. In this context, G3C3at is a sequent calculus that incorporates the rule of atomic excluded middle (at-ex-middle) in place of the rule of excluded middle (ex-middle). The admissibility of (ex-middle) in G3C3at was also demonstrated by them. Consequently, the cut-elimination theorem for a G3-style sequent calculus, referred to as G3C3, which is obtained from G3C3at by replacing (at-ex-middle) with (ex-middle), was also demonstrated by them in \cite{NW-JPL-2023}. Additionally, the first-order extensions of G3C, G3C3at, and G3C3 were also introduced and investigated by them. The systems G3C, G3C3at, and G3C3 were also used by Niki and Wansing in \cite{NW-JPL-2023} to explore the provable contradictions of C and C3.

%%%%%%%%%%% 2024/7/15 %%%%%%%%%%%%%
Several sequent calculi for some intermediate logics between C and C3 have recently been studied by Niki in \cite{Niki-FLAP-2024}. A three-sided sequent calculus for CN, under the name CC/TTm, has recently been introduced and investigated by \'Egr\'e et al. in \cite{ERS-2021}. 
A natural deduction system, NC2, and a two-sorted typed $\lambda$-calculus, 2$\lambda$, were introduced and investigated by Wansing in \cite{Wansing-IfColog-2016} for the bi-connexive propositional logic 2C. Natural deduction systems for two variants of connexive logics concerning non-classical interpretations of a certain kind between negation and implication were studied by Francez in \cite{Francez-2016}. 
%%%
In addition, some extensions of C were studied by Olkhovikov in \cite{Olkhovikov-FLAP-2016a,Olkhovikov-FLAP-2016b} and by Omori in \cite{Omori-2016}, although these studies are not concerned with sequent calculus or natural deduction system. 
%%%%%%%%%%%%%%%%%%%%%%%%%%%%%%%%%%%

The structure of this paper is as follows.
%%%
In Section \ref{c3-sequent-calculi-section}, we introduce Gentzen-style sequent calculi sC, sC3, sMC, and sCN for C, C3, MC, and CN, respectively. Additionally, we prove the cut-elimination theorems for these calculi. The calculi sC3, sMC, and sCN are obtained from sC by adding the excluded middle rule (ex-middle), the Peirce rule (Peirce), and both (ex-middle) and (Peirce), respectively. Moreover, we introduce alternative Gentzen-style sequent calculi sMC$^*$ and sCN$^*$ for MC and CN, respectively. These calculi are obtained from sC by adding the generalized excluded middle rule (g-ex-middle) and both (ex-middle) and (g-ex-middle), respectively. We then obtain a theorem establishing cut-free equivalence between sMC$^*$ (sCN$^*$) and sMC (sCN, resp.), along with presenting the cut-elimination theorem for sMC$^*$ and sCN$^*$.
%%%%
In Section \ref{c3-natural-deduction-section}, we introduce Gentzen-style natural deduction systems nC, nC3, nMC, and nCN for C, C3, MC, and CN, respectively. Additionally, we prove a theorem establishing equivalence between nC, (nC3, nMC, and nCN), and sC, (sC3, sMC$^*$, and nCN$^*$, resp.). Furthermore, we prove the normalization theorems for nC, nC3, nMC, and nCN.

%%%%%%%%%%%%%%%%%%%%%%%%%%%%%%%%%%%%%%%%%%%
\section{Gentzen-style sequent calculi}
\label{c3-sequent-calculi-section}

{\em Formulas} of connexive logics 
%%%%$ If this reference is cited in Introduction, then it redundant %%%%%%
\cite{ANGELL1962,MCCALL1966,WANSINGAIML2005,WANSING-CN} 
%%%%%%%%%%%%%%%%%%%%%%%%%%%%%%%%%%%%%%%%%%%%%%%%%%%%%%%%%%%%%%
are constructed using countably many propositional variables, the logical connectives $\LAND$ (conjunction), $\LOR$ (disjunction), $\I$ (implication), and $\NEG$ (connexive negation). We use small letters $p, q, ...$ to denote propositional variables, Greek small letters $\al, \be,...$ to denote formulas, and Greek capital letters $\GA, \DE,...$ to denote finite (possibly empty) 
% multisets 
sets of formulas. A {\em sequent} is an expression of the form \SEQ{\GA}{\ga}.  
%where $\ga$ is a formula or the empty multiset. 
We use the expression $L \vdash S$ to represent the fact that a sequent $S$ is provable in a sequent calculus $L$. We say that ``a rule $R$ of inference is {\em admissible} in a sequent calculus $L$'' if the following condition is satisfied: For any instance
{\tiny
$
\infer[]{S}{S_1 \cdots S_n}
$}of $R$, if $L \vdash S_i$ for all $i$, then $L \vdash S$. Furthermore, we say that ``$R$ is {\em derivable} in $L$'' if there is a derivation from $S_1, \cdots, S_n$ to $S$ in $L$.

We introduce Gentzen-style sequent calculi LJ$^+$ \cite{GENTZEN}, sC \cite{WANSINGAIML2005}, sC3, sMC, and sCN for positive intuitionistic logic, C \cite{WANSINGAIML2005}, C3 \cite{OW-AiML-2020}, MC \cite{WANSING-CN}, and CN \cite{C-NDJFL-2008}, respectively.

\begin{df}[LJ$^+$, sC, sC3, sMC, and sCN]
\label{sC-family-definition}~
\begin{enumerate}
\item 
{\rm LJ$^+$} is defined by the initial sequents and structural and logical inference rules of the following form, for any propositional variable $p$: 
{\footnotesize
$$
\SEQ{p, \GA}{p}~({\rm init1})
\quad
\infer[{\rm (cut)}]{\SEQ{\GA,\SI}{\ga}}{
  \SEQ{\GA}{\al}
  &
  \SEQ{\al, \SI}{\ga}
}
%\quad
%\infer[{\rm (we)}]{\SEQ{\al, \GA}{\ga}}{
% \SEQ{\GA}{\ga}
%}
$$
$$
\infer[{\rm (\I left)}]{\SEQ{\al\I\be, \GA,\DE}{\ga}}{
 \SEQ{\GA}{\al}
  &
 \SEQ{\be, \DE}{\ga}
}
\quad
\infer[{\rm (\I right)}]{\SEQ{\GA}{\al\I\be}}{
  \SEQ{\al, \GA}{\be}
}
\quad
\infer[{\rm (\LAND left)}]{\SEQ{\al\LAND\be, \GA}{\ga}}{
   \SEQ{\al, \be, \GA}{\ga}
}
\quad
\infer[{\rm (\LAND right)}]{\SEQ{\GA}{\al\LAND\be}}{
   \SEQ{\GA}{\al}
   &
   \SEQ{\GA}{\be}
}
$$
$$
\infer[{\rm (\LOR left)}]{\SEQ{\al\LOR\be, \GA}{\ga}}{
   \SEQ{\al, \GA}{\ga}
   &
   \SEQ{\be, \GA}{\ga}
}
\quad
\infer[{\rm (\LOR right1)}]{\SEQ{\GA}{\al\LOR\be}}{
   \SEQ{\GA}{\al}
}
\quad
\infer[{\rm (\LOR right2).}]{\SEQ{\GA}{\al\LOR\be}}{
   \SEQ{\GA}{\be}
}
$$
}
%%%%%%%%%%%%%
\item 
{\rm sC} is obtained from {\rm LJ$^+$} by adding the initial sequents and logical inference rules of the form:
{\footnotesize
$$
\SEQ{\NEG p, \GA}{\NEG p}~({\rm init2})
\quad
\infer[{\rm (\NEG left)}]{\SEQ{\NEG\NEG\al, \GA}{\ga}}{
   \SEQ{\al, \GA}{\ga}
}
\quad
\infer[{\rm (\NEG right)}]{\SEQ{\GA}{\NEG\NEG\al}}{
   \SEQ{\GA}{\al}
}
$$
$$
\infer[{\rm (\NEG\I left)}]{\SEQ{\NEG(\al\I\be), \GA,\DE}{\ga}}{
 \SEQ{\GA}{\al}
  &
 \SEQ{\NEG\be, \DE}{\ga}
}
\quad
\infer[{\rm (\NEG\I right)}]{\SEQ{\GA}{\NEG(\al\I\be)}}{
  \SEQ{\al, \GA}{\NEG\be}
}
$$
$$
\infer[{\rm (\NEG\LAND left)}]{\SEQ{\NEG(\al\LAND\be), \GA}{\ga}}{
   \SEQ{\NEG\al, \GA}{\ga}
   &
   \SEQ{\NEG\be, \GA}{\ga}
}
\quad
\infer[{\rm (\NEG\LAND right1)}]{\SEQ{\GA}{\NEG(\al\LAND\be)}}{
   \SEQ{\GA}{\NEG\al}
}
\quad
\infer[{\rm (\NEG\LAND right2)}]{\SEQ{\GA}{\NEG(\al\LAND\be)}}{
   \SEQ{\GA}{\NEG\be}
}
$$
$$
\infer[{\rm (\NEG\LOR left)}]{\SEQ{\NEG(\al\LOR\be), \GA}{\ga}}{
   \SEQ{\NEG\al, \NEG\be, \GA}{\ga}
}
\quad
\infer[{\rm (\NEG\LOR right)}.]{\SEQ{\GA}{\NEG(\al\LOR\be)}}{
   \SEQ{\GA}{\NEG\al}
   &
   \SEQ{\GA}{\NEG\be}
}
$$
}
%%%%%%
\item
{\rm sC3} and {\rm sMC} are obtained from {\rm sC} by adding the following excluded middle rule and Peirce rule, respectively: 
{\footnotesize
$$
\infer[(\mbox{\rm ex-middle})]{\SEQ{\GA}{\ga}}{
  \SEQ{\NEG\al,\GA}{\ga} 
   &
  \SEQ{\al, \GA}{\ga}
}
\quad
\infer[({\rm Peirce}).]{\SEQ{\GA}{\al}}{
  \SEQ{\al\I\be, \GA}{\al} 
}
$$}
%%%%%
%\item
%{\rm sMC} is obtained from {\rm sC} by adding the Peirce rule of the form:
%{\footnotesize
%$$
%\infer[({\rm Peirce}).]{\SEQ{\GA}{\al}}{
%  \SEQ{\al\I\be, \GA}{\al} 
%}
%$$}
%%%%%%%
\item
{\rm sCN} is obtained from {\rm sC3} by adding {\rm (Peirce)}. 
\end{enumerate}
\end{df}

\begin{rmk}~
\begin{enumerate}
%%%%%%%
\item
It is known that single-succedent Gentzen-style sequent calculi for classical logic are obtained from Gentzen's sequent calculus {\rm LJ} (or other variants such as the G3-style sequent calculus {\rm G3ip}) for intuitionistic logic by adding one of {\rm (ex-middle)}, {\rm (Peirce)}, and their variants. These single-succeddent calculi have been studied by several researchers {\rm \cite{Curry-1963,Felscher-1975,Gordeev-1987,Africk-1992,Von-Plato-draft-1998,NP-2001,Kamide-BSL-2005,KAMIDE-DRAFT-2022}}. For a survey on these calculi, see, for example, {\rm \cite{Kamide-BSL-2005,KAMIDE-DRAFT-2022}}.

%%%%%%%%
\item
{\rm (ex-middle)}, which corresponds to the law of excluded middle $\NEG\al\LOR\al$, was introduced and investigated by von Plato {\rm \cite{Von-Plato-draft-1998,NP-2001}}, although the name {\rm (ex-middle)} was not used by him. He showed that {\rm (ex-middle)} can be restricted to the inference rule of the form: 
{\footnotesize
$$
\infer[(\mbox{\rm at-ex-middle})]{\SEQ{\GA}{\ga}}{
  \SEQ{\NEG p, \GA}{\ga} 
   &
  \SEQ{p, \GA}{\ga}
}
$$}where $p$ is a propositional variable. Namely, {\rm (at-ex-middle)} and {\rm (ex-middle)} are equivalent over intuitionistic logic. He proved the cut-elimination theorems for some sequent calculi with {\rm (at-ex-middle)} or {\rm (ex-middle)}.

%%%%%%%%
\item
{\rm (Peirce)}, which corresponds to the Peirce law $((\al\I\be)\I\al)\I\al$, was introduced and investigated by Curry {\rm \cite{Curry-1963}}, Felscher {\rm \cite{Felscher-1975}}, Gordeev {\rm \cite{Gordeev-1987}}, and Africk {\rm \cite{Africk-1992}}. The cut-elimination theorem for {\rm LJ} $+$ {\rm (Peirce)} was proved by them. Specifically, Africk  {\rm \cite{Africk-1992}} obtained a simple embedding-based proof of the cut-elimination theorem for {\rm LJ} $+$ {\rm (Peirce)}. The subformula property for a version of {\rm LJ} $+$ {\rm (Peirce)} without the falsity constant $\bot$ was shown by Gordeev. Specifically, he proved in {\rm \cite{Gordeev-1987}} that $\be$ in {\rm (Peirce)} can be restricted to a subformula of some formulas in $(\GA, \al)$.

%%%%%%%%%
\item 
Gentzen's {\rm LK} for classical logic, {\rm LJ} $+$ {\rm (ex-middle)}, and {\rm LJ} $+$ {\rm (Peirce)} are theorem-equivalent within the language $\{\LAND, \LOR, \I, \neg, \bot\}$. However, {\rm sC3}, {\rm sMC}, and {\rm sCN} (and their corresponding logics {\rm C3}, {\rm MC}, and {\rm CN}) are not logically-equivalent. This fact will be shown in Theorem \ref{separation-theorem}. 
%%%%%%%%%  It was moved to Introduction. %%%%%%%%%% 
%\item
%The cut-elimination theorem for {\rm sC} was originally proved by Wansing in {\rm \cite{WANSINGAIML2005}}, although the name {\rm sC} was not used by him. The cut-elimination theorems for G3-style sequent calculi {\rm G3C} and {\rm G3C3at} for {\rm C} and {\rm C3}, respectively, were originally proved by Omori and Wansing in {\rm \cite{OW-AiML-2020}}, where {\rm G3C3at} is a sequent calculus having {\rm (at-ex-middle)} instead of {\rm (ex-middle)}. The admissibility of {\rm (ex-middle)} in {\rm G3C3at} was also shown by them. Thus, the cut-elimination theorem for a G3-style sequent calculus {\rm G3C3}, which is obtained from {\rm G3C3at} by replacing {\rm (at-ex-middle)} with {\rm (ex-middle)}, was also shown by them in {\rm \cite{NW-JPL-2023}}. The first-order extensions of {\rm G3C}, {\rm G3C3at}, and {\rm G3C3} were also introduced and investigated by them.The systems  {\rm G3C}, {\rm G3C3at}, and {\rm G3C3} were also used by Niki and Wansing in {\rm \cite{NW-JPL-2023}} for investigating the provable contradictions of {\rm C} and {\rm C3}.
\end{enumerate}
\end{rmk}

%We have the following proposition.

\begin{prop}
\label{initial-sequent-prop}
Let $L$ be {\rm LJ$^+$}, {\rm sC}, {\rm sC3}, {\rm sMC}, or {\rm sCN}. 
For any formula $\al$ and any set $\GA$ of formulas, we have: $L$ $\vdash$ \SEQ{\al, \GA}{\al}.
\end{prop}
\PROOF
By induction on $\al$. 
\QED
%\\

\begin{prop}
\label{we-prop}
Let $L$ be {\rm LJ$^+$}, {\rm sC}, {\rm sC3}, {\rm sMC}, or {\rm sCN}. 
The following rule is admissible in cut-free $L$:
$$
\infer[{\rm (we)}.]{\SEQ{\al, \GA}{\ga}}{
 \SEQ{\GA}{\ga}
}
$$
\end{prop}
\PROOF
By induction on the proofs $P$ of \SEQ{\GA}{\ga} of (we) in cut-free $L$.
\QED
\\

The following cut-elimination theorems for LJ$^+$ and sC are well-known.

\begin{thm}[Cut-elimination for {\rm LJ$^+$} and {\rm sC} \cite{GENTZEN,WANSINGAIML2005}]
\label{cut-eli-sC}
Let  $L$ be {\rm LJ$^+$} or {\rm sC}. 
The rule {\rm (cut)} is admissible in cut-free $L$.
\end{thm}

We now show the cut-elimination theorems for sC3, sMC, and sCN.

\begin{thm}[Cut-elimination for {\rm sC3}, {\rm sMC}, and {\rm sCN}]
\label{cut-eli-C-systems}
Let  $L$ be {\rm sC3}, {\rm sMC}, or {\rm sCN}. 
The rule {\rm (cut)} is admissible in cut-free $L$.
\end{thm}
\PROOF
(Sketch).
We give a sketch of the proof.

%%%%%%%
$\bullet$
First, we show the cut-elimination theorem for sC3. 
It is known that the cut-elimination theorem for the G3-style sequent calculus G3C3 for C3, which has (ex-middle), holds {\rm \cite{OW-AiML-2020}}. Then, we can show the cut-free equivalence between G3C3 and sC3. Thus, from this equivalence and the cut-elimination theorem for G3C3, we obtain the cut-elimination theorem for sC3.

%%%%%%%
$\bullet$
Second, we show the cut-elimination theorem for sMC. 
It is known that the cut-elimination theorem for LJ + (Peirce) holds. This theorem was proved directly and indirectly by using the methods by Gordeev \cite{Gordeev-1987} and Africk \cite{Africk-1992}. Thus, the cut-elimination theorem for the negation-less fragment (i.e., LJ$^+$ + (Peirce)) of LJ + (Peirce) holds because LJ + (Peirce) is a conservative extension of LJ$^+$ + (Peirce) by the cut-elimination theorem for LJ + (Peirce). Then, we can show a theorem for embedding (cut-free) sMC into (cut-free) LJ$^+$ + (Peirce), and by using this theorem, we can show the cut-elimination theorem for sMC. We will show this in the following.

%%%%%%%
Prior to showing the embedding theorem, we introduce a translation of sMC to LJ$^+$ + (Peirce). 
Let $\PH$ be a set of propositional variables and $\PH'$ be the set $\{p'~|~p \in \PH \}$ of propositional variables. Then, the language ${\cal L}_{\rm MC}$ of sMC is defined using $\PH$, $\LAND$, $\LOR$, $\I$, and $\NEG$. The language ${\cal L}_{\rm Int+}$ of LJ$^+$ is obtained from ${\cal L}_{\rm MC}$ by replacing $\NEG$ with $\PH'$. 
A mapping $f$ from ${\cal L}_{\rm MC}$ to ${\cal L}_{\rm Int+}$ is defined inductively by:
%\begin{enumerate} 
%\item
(1) for any $p \in \PH$, 
$f(p) := p$ and $f(\NEG p) := p' \in \PH'$; 
%\item
(2) $f(\al~\sharp~\be) := f(\al)~\sharp~f(\be)$ with $\sharp \in \{\LAND, \LOR, \I\}$; 
%\item 
(3) $f(\NEG\NEG\al) := f(\al)$; 
%\item
(4) $f(\NEG (\al \LAND \be)) := f(\NEG\al) \LOR f(\NEG\be)$; 
%\item
(5) $f(\NEG (\al \LOR \be)) := f(\NEG\al) \LAND f(\NEG\be)$; and 
%\item
(6) $f(\NEG (\al \I \be)) := f(\al) \I f(\NEG\be)$. 
%\end{enumerate}
%%%%%
An expression $f(\GA)$ denotes the result of replacing every occurrence of a formula $\al$ in $\GA$ by an occurrence of $f(\al)$. 
%Analogous notion is used for the other mappings discussed later. 
We remark that a similar translation defined as above has been used by Gurevich {\rm  \cite{GUREVICH}}, Rautenberg {\rm \cite{RAUTENBERG}} and Vorob'ev {\rm \cite{VOR52}} to embed Nelson's constructive logic {\rm \cite{AN84,NELSON}} into positive intuitionistic logic.

We then obtain the following theorem for embedding sMC into LJ$^+$ $+$ (Peirce): 
\begin{enumerate}
\item 
{\rm sMC} $\vdash$  $\SEQ{\GA}{\ga}$ iff~ 
{\rm LJ$^+$} $+$ {\rm (Peirce)}  $\vdash$ $\SEQ{f(\GA)}{f(\ga)}$,  

\item
{\rm sMC} $-$ {\rm (cut)} $\vdash$  $\SEQ{\GA}{\ga}$ iff~ 
{\rm LJ$^+$} $+$ {\rm (Peirce)} $-$ {\rm (cut)} $\vdash$ $\SEQ{f(\GA)}{f(\ga)}$. 
\end{enumerate}  
The proof of this theorem is almost the same as that for the theorem for embedding sC or a Gentzen-style sequent calculus for Nelson's paraconsistent four-valued logic N4 into LJ$^+$. For more information on these embedding theorems, see, for example,  \cite{KWTCS2012,KW-BOOK-2015,KWBOOK2011,KWIFCOLOG2016,Kamide-JPL-2022-2}.

We are ready to prove of the cut-elimination theorem for sMC. 
Suppose that {\rm sMC} $\vdash$ $\SEQ{\GA}{\ga}$.  
Then, we have {\rm LJ$^+$} $+$ {\rm (Peirce)} $\vdash$ $\SEQ{f(\GA)}{f(\ga)}$ by 
the statement (1) of the theorem, and hence 
{\rm LJ$^+$} $+$ {\rm (Peirce)} $-$ (cut) $\vdash$ $\SEQ{f(\GA)}{f(\ga)}$ by   
the cut-elimination theorem for {\rm LJ$^+$} $+$ {\rm (Peirce)}.  
Then, by the statement (2) of the theorem, we obtain 
{\rm sMC} $-$ (cut) $\vdash$ $\SEQ{\GA}{\ga}$.

%%%%%%%
$\bullet$
Finally, the cut-elimination theorem for sCN can be proved in a similar way as for sMC. 
%\end{enumerate}
\QED
%\\

\begin{thm}[Separation of {\rm C}, {\rm C3}, {\rm MC}, and {\rm CN}]
\label{separation-theorem}
The logics {\rm C}, {\rm C3}, {\rm MC}, and {\rm CN} are not logically-equivalent.
\end{thm}
\PROOF
By Theorem \ref{cut-eli-C-systems}. 
%
%To show this, we use {\rm sC}, {\rm sC3}, {\rm sMC}, {\rm sCN}, and Theorem \ref{cut-eli-C-systems}. 
%Let $p$ and $q$ be distinct propositional variables. 
%Then, we consider only the following facts: 
%\begin{enumerate}
%\item
%\SEQ{}{((p\I q)\I p)\I p} is provable in cut-free {\rm sMC}, but not provable in cut-free {\rm sC3},  
%
%\item
%\SEQ{}{\NEG p \LOR p} is provable in cut-free {\rm sC3}, but not provable in cut-free {\rm sMC}. 
%\end{enumerate}
%The unprovabilities of these sequents are guaranteed by Theorem \ref{cut-eli-C-systems}. 
%We thus show only the case for {\rm sMC} $-$ (cut) $\vdash$ \SEQ{}{((p\I q)\I p)\I p} by: 
%{\footnotesize
%$$
%\infer[(\I {\rm left}).]{\SEQ{}{((p\I q)\I p)\I p}}{
%     \infer[({\rm Peirce})]{\SEQ{(p\I q)\I p}{p}}{
%           \infer[(\I {\rm left})]{\SEQ{p\I q, (p\I q)\I p}{p}}{
%                 \infer[(\I {\rm right})]{\SEQ{p\I q}{p \I q}}{
%                       \infer[(\I {\rm left})]{\SEQ{p\I q, p}{q}}{
%                            \SEQ{p}{p}~{\rm (init1)}
%                             &
%                            \SEQ{q}{q}~{\rm (init1)}
%                       }
%                 }
%                 &
%                 \SEQ{p}{p}~{\rm (init1)}
%           }
%     }
%}
%$$}
\QED
\\

Next, we introduce alternative Gentzen-style sequent calculi sMC$^*$ and sCN$^*$ for MC and CN, respectively. These calculi will be used to prove the normalization theorems for the natural deduction systems nMC and nCN for MC and CN, respectively.

\begin{df}[sMC$^*$ and sCN$^*$]
\label{sMC-star-sCN-star-def}~
\begin{enumerate}
\item
{\rm sMC$^*$} is obtained from {\rm sC} by adding the generalized excluded middle rule of the form:
{\footnotesize
$$
\infer[(\mbox{\rm g-ex-middle}).]{\SEQ{\GA}{\ga}}{
  \SEQ{\al\I\be,\GA}{\ga} 
   &
  \SEQ{\al, \GA}{\ga}
}
$$
}
\item
{\rm sCN$^*$} is obtained from {\rm sC3} by adding {\rm (g-ex-middle)}. 
\end{enumerate}
\end{df}

\begin{rmk}~
\begin{enumerate}
%%%%%%%%
\item
{\rm (g-ex-middle)}, which corresponds to the generalized law of excluded middle $(\al\I\be)\LOR\al$, was introduced and investigated by Kamide in {\rm \cite{Kamide-BSL-2005}}, although the name {\rm (g-ex-middle)} was not used by him. He proved the cut-elimination theorem for {\rm LJ} $+$ {\rm (g-ex-middle)} using the method by Africk {\rm \cite{Africk-1992}}. 

%%%%%%%%
\item 
{\rm LJ} $+$ {\rm (g-ex-middle)} is regarded as a sequent calculus for classical logic. Actually, {\rm (g-ex-middle)} and {\rm (ex-middle)} are equivalent over positive intuitionistic logic. 
{\rm (g-ex-middle)} is regarded as a generalization of {\rm (ex-middle)} if we assume the falsity constant $\bot$ and the definition $\NEG\al := \al\I\bot$. {\rm (g-ex-middle)} is also regarded as a generalization of {\rm (Peirce)} and it was referred to as generalized Peirce rule (named {\rm (g-Peirce)}) in {\rm \cite{Kamide-BSL-2005}}.
%%
%%%%%%%%%
%\item
%The following is an example proof of \SEQ{}{(p\I q)\LOR p} in cut-free {\rm sMC$^*$}:
%$$
%\infer[\mbox{\rm (g-ex-middle)}.]{\SEQ{}{(p\I q)\LOR p}}{
%     \infer[(\LOR {\rm right1})]{\SEQ{p\I q}{(p\I q)\LOR p}}{
%          \infer[(\I {\rm right})]{\SEQ{p\I q}{p\I q}}{
%               \infer[(\I {\rm left})]{\SEQ{p\I q, p}{q}}{
%                    \SEQ{p}{p}~{\rm (init1)}
%                    &
%                    \SEQ{q}{q}~{\rm (init1)}
%               }
%          }
%     }
%     &
%     \infer[(\LOR {\rm right2})]{\SEQ{p}{(p\I q)\LOR p}}{
%           \SEQ{p}{p}~{\rm (init1)}
%     }
%}
%$$
\end{enumerate}
\end{rmk}

%We have the following proposition.

\begin{prop}
Let $L$ be {\rm sMC$^*$} or {\rm sCN$^*$}. 
For any formula $\al$ and set $\GA$ of formulas, we have: $L$ $\vdash$ \SEQ{\al, \GA}{\al}.
\end{prop}
\PROOF
By induction on $\al$. 
\QED
%\\

\begin{prop}
\label{we-prop-2}
Let $L$ be {\rm sMC$^*$} or {\rm sCN$^*$}. 
The rule {\rm (we)} is admissible in cut-free $L$. 
%$$
%\infer[{\rm (we)}.]{\SEQ{\al, \GA}{\ga}}{
% \SEQ{\GA}{\ga}
%}
%$$
\end{prop}
\PROOF
Similar to the proof of Proposition \ref{we-prop}. 
%By induction on the proofs $P$ of \SEQ{\GA}{\ga} of (we) in cut-free $L$.
\QED
%\\

%We obtain the following theorem. 

\begin{thm}[Equivalence between sMC (sCN) and sMC$^*$ (sCN$^*$)]
\label{equivalence-sMC and sMC-star}
Let $L_1$ and $L_2$ be the sequent calculi {\rm sMC} and {\rm sCN}, respectively. 
Let $L_1^*$ and $L_2^*$ be the sequent calculi {\rm sMC$^*$} and {\rm sCN$^*$}, respectively. 
For any $i \in \{1, 2\}$, we have:
\begin{enumerate}
\item 
$L_i$ $\vdash$ \SEQ{\GA}{\ga}~iff~ 
$L_i^*$ $\vdash$ \SEQ{\GA}{\ga},

\item 
$L_i$ $-$ {\rm (cut)} $\vdash$ \SEQ{\GA}{\ga}~iff~ 
$L_i^*$ $-$ {\rm (cut)} $\vdash$ \SEQ{\GA}{\ga}.
\end{enumerate}
\end{thm}
\PROOF
Straightforward. 
%We show only (2). 
%The fact that $L_i$ $-$ {\rm (cut)} $\vdash$ \SEQ{\GA}{\ga} implies $L_i^*$ $-$ {\rm (cut)} $\vdash$ \SEQ{\GA}{\ga} is obvious because (Peirce) is an instance of (g-ex-middle). Thus, we show that $L_i^*$ $-$ {\rm (cut)} $\vdash$ \SEQ{\GA}{\ga} implies $L_i$ $-$ {\rm (cut)} $\vdash$ \SEQ{\GA}{\ga} by induction on the proofs $P$ of \SEQ{\GA}{\ga} in $L_i^*$ $-$ {\rm (cut)}. We distinguish the cases according to the last inference of $P$ and show only the following case. 
%%%%%%%%%%%%%%%%%%%
%Case (g-ex-middle): 
%The last inference of $P$ is of the form: 
%{\footnotesize
%$$
%\infer[(\mbox{\rm g-ex-middle}).]{\SEQ{\GA}{\ga}}{
%   \infer*[]{\SEQ{\al\I\be,\GA}{\ga}}{
%   } 
%   &
%   \infer*[]{\SEQ{\al, \GA}{\ga}}{
%   }
%}
%$$}By induction hypotheses, we have: 
%$L_i$ $-$ {\rm (cut)} $\vdash$ \SEQ{\al\I\be, \GA}{\ga} and 
%$L_i$ $-$ {\rm (cut)} $\vdash$ \SEQ{\al, \GA}{\ga}.
%Then, we obtain the required fact: 
%{\footnotesize
%$$
%\infer[{\rm (Peirce)}.]{\SEQ{\al\I\be, \GA}{\ga}}{
%         \infer[(\I {\rm left})]{\SEQ{\ga\I\al, \al\I\be, \GA}{\ga}}{
%               \infer*[Ind. hyp.]{\SEQ{\al\I\be, \GA}{\ga}}{
%               }
%               &
%               \infer*[Ind. hyp.]{\SEQ{\al, \GA}{\ga}}{
%               }
%         }
%}
%$$
%}
\QED
%\\

%We then  obtain the following cut-elimination theorems for sMC$^*$ and sCN$^*$. 

\begin{thm}[Cut-elimination for {\rm sMC$^*$} and {\rm sCN$^*$}]
\label{cut-eli-sMC-star}
Let  $L$ be {\rm sMC$^*$} or {\rm sCN$^*$}. 
The rule {\rm (cut)} is admissible in cut-free $L$.
\end{thm}
\PROOF
By Theorems \ref{cut-eli-C-systems} and \ref{equivalence-sMC and sMC-star}.
\QED

%%%%%%%%%%%%%%%%%%%%%%%%%%%%%%%%%%%%%%%%%%%
\section{Gentzen-style natural deduction systems}
\label{c3-natural-deduction-section}

We now define Gentzen-style natural deduction systems NJ$^+$, nC, nC3, nMC, and nCN
for positive intuitionistic logic, C, C3, MC, and CN, respectively. 
%For general information on Gentzen-style natural deduction system, see \cite{GENTZEN,PRAWITZ}.
%%%%% 
We use the notation $[\al]$ in the definitions of natural deduction systems to denote the discharged assumption (i.e., the formula $\al$ is a discharged assumption by the underlying logical inference rule).
%%%%%%
%An expression $[\al]^*$ denotes at least one discharge of $\al$.

\begin{df}[NJ$^+$, nC, nC3, nMC, and nCN]
\label{natural-deduction-definition}~
%An expression $[\al]^*$ denotes at least one discharge of $\al$.
\begin{enumerate}
%%%%%
\item
{\rm NJ$^+$} is defined as the logical inference rules of the form, where in {\rm ($\I$I)} the discharge can be vacuous:
{\footnotesize
$$
\infer[(\I {\rm I})]{\al\I\be}{
  \infer*[]{\be}{
    [\al]
  }
}
\quad
\infer[(\I {\rm E})]{\be}{
  \al\I\be
  &
  \al
}
\quad
\infer[(\LAND {\rm I})]{\al\LAND\be}{
\al
&
\be
}
\quad
\infer[(\LAND {\rm E1})]{\al}{
 \al\LAND\be
}
\quad
\infer[(\LAND {\rm E2})]{\be}{
 \al\LAND\be
}
$$
$$
\infer[(\LOR {\rm I1})]{\al\LOR \be}{
 \al
}
\quad
\infer[(\LOR {\rm I2})]{\al\LOR \be}{
 \be
}
\quad
\infer[(\LOR {\rm E}).]{\ga}{
  \al\LOR\be
  &
  \infer*[]{\ga}{
     [\al]
   }
   &
  \infer*[]{\ga}{
     [\be]
  }
}
$$
}
%%%%%%%%
\item
{\rm nC} is obtained from {\rm NJ$^+$} 
by 
%replacing {\rm ($\I$I)} with the following restricted versions of {\rm ($\I$I)}:
%{\footnotesize
%$$
%\infer[{\rm (Wk)}]{\al\I\be}{
%  \be
%}
%\quad
%\infer[(\I{\rm I}^*)]{\al\I\be}{
%  \infer*[]{\be}{
%    [\al]^*
%  }
%}
%$$}where $[\al]^*$ represents at least one discharge of $\al$, and 
adding the negated logical inference rules of the form:
{\footnotesize
$$
\infer[(\NEG\NEG {\rm I})]{\NEG\NEG\al}{
 \al
}
\quad
\infer[(\NEG\NEG {\rm E})]{\al}{
  \NEG\NEG\al
}
\quad
%\infer[{\rm (\NEG Wk)}]{\NEG(\al\I\be)}{
%  \NEG\be
%}
%\quad
\infer[(\NEG\I {\rm I})]{\NEG(\al\I\be)}{
  \infer*[]{\NEG\be}{
    [\al]
  }
}
\quad
\infer[(\NEG\I {\rm E})]{\NEG\be}{
  \NEG(\al\I\be)
  &
  \al
}
$$
$$
\infer[(\NEG\LAND {\rm I1})]{\NEG(\al\LAND\be)}{ 
   \NEG\al 
}
\quad
\infer[(\NEG\LAND {\rm I2})]{\NEG(\al\LAND\be)}{ 
   \NEG\be 
}
\quad
\infer[(\NEG\LAND {\rm E})]{\ga}{ 
   \NEG(\al\LAND\be) 
   &
   \infer*[]{\ga}{
       [ \NEG\al ] 
    }
   &
   \infer*[]{\ga}{
       [ \NEG\be ]
   }
}
$$
$$
\infer[(\NEG\LOR {\rm I})]{\NEG(\al\LOR\be)}{ 
   \NEG\al
   &
   \NEG\be
}
\quad
\infer[(\NEG\LOR {\rm E1})]{\NEG\al}{ 
   \NEG(\al\LOR\be) 
}
\quad
\infer[(\NEG\LOR {\rm E2}).]{\NEG\be}{ 
   \NEG(\al\LOR\be) 
}
$$
}
%%%%%%%%
\item
{\rm nC3} and {\rm nMC} are obtained from {\rm nC} by adding the following excluded middle rule and generalized excluded middle rule, respectively: 
{\footnotesize
$$
\infer[({\rm EM})]{\ga}{
  \infer*[]{\ga}{
    [\NEG\al]
  }
  &
  \infer*[]{\ga}{
   [\al]
  }
}
\quad
\infer[({\rm GEM}).]{\ga}{
  \infer*[]{\ga}{
    [\al\I\be]
  }
  &
  \infer*[]{\ga}{
   [\al]
  }
}
$$
}
%%%%%%%%
%\item
%{\rm nMC} is obtained from {\rm nC3} by adding the generalized rule of excluded middle of the form:
%{\footnotesize
%$$
%\infer[({\rm GEM}).]{\ga}{
%  \infer*[]{\ga}{
%    [\al\I\be]
%  }
%  &
%  \infer*[]{\ga}{
%   [\al]
%  }
%}
%$$
%}
%%%%%%%%
\item
{\rm nCN} is obtained from {\rm nC3} by adding {\rm (GEM)}. 
\end{enumerate}
\end{df}

\begin{rmk}
%We make the following remarks. 
%\begin{enumerate}
%%%%%%%%%%
%\item
{\rm (EM)} and its restricted version {\rm (EM-at)} with the propositional variable discharged assumptions were originally introduced by von Plato {\rm \cite{Von-Plato-draft-1998}}, and called there {\em Gem}  and {\em Gem-at}, respectively. He proved normalization theorems for systems with {\rm (EM)} or {\rm (EM-at)}. 
\end{rmk}

Next, we define some notions for the natural deduction systems.

%%%%%%% 
\begin{df}
{\rm ($\I$I)}, {\rm ($\LAND$I)}, {\rm ($\LOR$I1)}, {\rm ($\LOR$I2)}, 
{\rm ($\NEG\NEG$I)}, {\rm ($\NEG\I$I)}, {\rm ($\NEG\LAND$I1)}, {\rm ($\NEG\LAND$I2)}, {\rm ($\NEG\LOR$I)}, {\rm (EM)}, and {\rm (GEM)} are called {\em introduction rules}, and {\rm ($\I\,$E)}, {\rm ($\LAND$E1)}, {\rm ($\LAND$E2)}, {\rm ($\LOR$E)}, {\rm ($\NEG\NEG$E)}, {\rm ($\NEG\I\,$E)}, {\rm ($\NEG\LAND$E)}, {\rm ($\NEG\LOR$E1)}, and {\rm ($\NEG\LOR$E2)} are called {\em elimination rules}. 
%%%%
The notions of {\em major} and {\rm minor premises} of the inference rules without {\rm (EM)} and {\rm (GEM)} are defined as usual. 
%%%%
The notions of {\em derivation}, {\em (open and discharged) assumptions} of derivation, and {\em end-formula} of derivation are also defined as usual. 
%%%%%%
Any derivation starts with an assumption $\al$ can be considered a derivation of $\al$ from itself.
%%%% Critical set --> multiset  %%%%%%%%%%%%%%%%%%%%%
For a derivation $\mathcal{D}$, we use the expression {\rm oa($\mathcal{D}$)} to denote the set of open assumptions of $\mathcal{D}$ and the expression {\rm end($\mathcal{D}$)} to denote the end-formula of $\mathcal{D}$. 
%%%%%%%%%%%%%%%%%%%%%%%%%%%%%%%%%%%%%%%%%%%%%%%
A formula $\al$ is said to be {\em provable} in a natural deduction system $N$ if there exists a derivation in $N$ with no open assumptions whose end-formula is $\al$. 
%%%%
%Similar terminologies and notions will also be used for other systems discussed in this study. 
\end{df}

\begin{rmk}
%%%
%Note that {\rm (Wk)} and {\rm ($\NEG$Wk)} are not treated as introduction rules. This is crucial for defining the notion of maximum formula. 
%%%%%%
There are no notions of major and minor premises of {\rm (EM)} and {\rm (GEM)}. 
Namely, both the premises of {\rm (EM)} and {\rm (GEM)} are neither major nor minor premise. 
%%%
In this study, {\rm (EM)} and {\rm (GEM)} are treated as introduction rules. 
\end{rmk}

Next, we define a reduction relation $\gg$ on the set of derivations in the natural deduction systems. Prior to defining $\gg$, we define some notions concerning $\gg$.

\begin{df}
Let $L$ be {\rm nC}, {\rm nC3}, {\rm nMC}, or {\rm nCN}. 
Let $\al$ be a formula occurring in a derivation $\mathcal{D}$ in $L$. 
Then, $\al$ is called a {\em maximum formula} in $\mathcal{D}$ if $\al$ satisfies the following conditions: 
\begin{enumerate}
\item
$\al$ is the conclusion of an introduction rule, {\rm ($\LOR$E)}, or {\rm ($\NEG\LAND$E)},

\item
$\al$ is the major premise of an elimination rule.
\end{enumerate}
A derivation is said to be {\em normal} if it contains no maximum formula. 
%%%%%
The notion of substitution of derivations to assumptions is defined as usual. We assume that the set of derivations is closed under substitution. 
\end{df}

\begin{df}[Reduction relation]
\label{nc-family-reduction}
Let $\ga$ be a maximum formula in a derivation that is the conclusion of an inference rule $R$. 

\begin{enumerate}
%%%%%%%%%%%%%%%%%%%%%%%%%%%%%%%%%%%%%%%%%
\item
The definition of the reduction relation $\gg$ at $\ga$ in {\rm nC} is obtained by the following conditions.
\begin{enumerate}
%%%%%%%%
\item
$R$ is {\rm ($\I$I)} and $\ga$ is $\al\I\be$: 
{\footnotesize
$$
\infer[(\I{\rm E})]
{\be}{
   \infer[(\I{\rm I})]
   {\al\I\be}{
      \infer*[{\mathcal D}]
      {\be}{
         [\al]
      }
   }
   &
   \infer*[{\mathcal E}]{\al}{
   }
}
\quad\quad \gg \quad\quad
  \infer*[{\mathcal D}]{\be.}{
     \infer*[{\mathcal E}]{\al}{
      }
  }
$$
}
%%%%%%%%%
\item
$R$ is {\rm ($\LAND$I)} and $\ga$ is $\al_1\LAND\al_2$: 
{\footnotesize
$$
\infer[(\LAND {\rm E}i)]{\al_i}{
   \infer[(\LAND {\rm I})]{\al_1\LAND\al_2}{
       \infer*[{\mathcal D}_1]{\al_1}{
        }
        &
       \infer*[{\mathcal D}_2]{\al_2}{
        }
   }  
}
\quad\quad \gg \quad\quad 
   \infer*[{\mathcal D}_i]{\al_i}{
   }
\quad 
where~i~is~1~or~2. 
$$}
%where $i$ is $1$ or $2$. 
%%%%%%%%%
\item
$R$ is {\rm ($\LOR$I1)} or {\rm ($\LOR$I2)} and $\ga$ is $\al_1\LOR\al_2$: 
{\footnotesize
$$
\infer[(\LOR {\rm E})]{\de}{
  \infer[(\LOR {\rm I}i)]{\al_1\LOR\al_2}{
      \infer*[{\mathcal D}]{\al_i}{
       }
  }
  &
  \infer*[{\mathcal E}_1]{\de}{
    [\al_1]
  }
  &
  \infer*[{\mathcal E}_2]{\de}{
    [\al_2]
  }
}
\quad\quad \gg \quad\quad 
  \infer*[{\mathcal E}_i]{\de}{
      \infer*[{\mathcal D}]{\al_i}{
      }
  }
\quad 
where~i~is~1~or~2. 
$$}
%where $i$ is $1$ or $2$. 
%%%%%%%
\item
$R$ is {\rm ($\LOR$E)}: 
{\scriptsize
$$
\infer[R']{\de}{
   \infer[(\LOR {\rm E})]{\ga}{
       \infer*[{\mathcal D}_1]{\al\LOR\be}{
        }
        &
       \infer*[{\mathcal D}_2]{\ga}{
          [\al]
        }
        &
       \infer*[{\mathcal D}_3]{\ga}{
          [\be]
        }
   }  
   &
   \infer*[{\mathcal E}_1]{\de_1}{
   }
   &
   \infer*[{\mathcal E}_2]{\de_2}{
   }
}
\quad\quad \gg \quad\quad 
\infer[(\LOR {\rm E})]{\de}{ 
   \infer*[{\mathcal D}_1]{\al\LOR\be}{
   }
   &
   \infer[R']{\de}{
      \infer*[{\mathcal D}_2]{\ga}{
          [\al]
       }
       &
       \infer*[{\mathcal E}_1]{\de_1}{
       }
       &
       \infer*[{\mathcal E}_2]{\de_2}{
       }
   }
   &
   \infer[R']{\de}{
      \infer*[{\mathcal D}_3]{\ga}{
          [\be]
       }
       &
       \infer*[{\mathcal E}_1]{\de_1}{
       }
       &
       \infer*[{\mathcal E}_2]{\de_2}{
       }
   }
}
$$}where $R'$ is an arbitrary inference rule, and both ${\mathcal E}_1$ and ${\mathcal E}_2$ are derivations of the minor premises of $R'$ if they exist.

%%%%%%%
\item 
$R$ is {\rm ($\NEG\NEG$I)}, and $\ga$ is $\NEG\NEG\al$: 
{\footnotesize
$$
\infer[(\NEG\NEG {\rm E})]{\al}{
   \infer[(\NEG\NEG {\rm I})]{\NEG\NEG\al}{
       \infer*[\mathcal{D}]{\al}{
        }
   }  
}
\quad\quad \gg \quad\quad 
   \infer*[\mathcal{D}]{\al.}{
   }
$$
}
%%%%%%%%% 
\item
$R$ is {\rm ($\NEG\I$I)} and $\ga$ is $\NEG(\al\I\be)$: 
{\footnotesize
$$
\infer[(\NEG\I{\rm E})]
{\NEG\be}{
   \infer[(\NEG\I{\rm I})]
   {\NEG(\al\I\be)}{
      \infer*[{\mathcal D}]
      {\NEG\be}{
         [\al]
      }
   }
   &
   \infer*[{\mathcal E}]{\al}{
   }
}
\quad\quad \gg \quad\quad
  \infer*[{\mathcal D}]{\NEG\be.}{
     \infer*[{\mathcal E}]{\al}{
      }
  }
$$
}
%%%%%%%%%
\item
$R$ is {\rm ($\NEG\LAND$I1)} or {\rm ($\NEG\LAND$I2)} and $\ga$ is $\NEG(\al_1\LAND\al_2)$: 
{\footnotesize
$$
\infer[(\NEG\LAND {\rm E})]{\de}{
  \infer[(\NEG\LAND {\rm I}1)]{\NEG(\al_1\LAND\al_2)}{
      \infer*[\mathcal{D}]{\NEG\al_i}{
       }
  }
  &
  \infer*[{\mathcal{E}}_1]{\de}{
    [\NEG\al_1]
  }
  &
  \infer*[{\mathcal{E}}_2]{\de}{
    [\NEG\al_2]
  }
}
\quad\quad \gg \quad\quad 
  \infer*[{\mathcal{E}}_i]{\de}{
      \infer*[\mathcal{D}]{\NEG\al_1}{
      }
  }
\quad 
where~i~is~1~or~2. 
$$}
%where $i$ is $1$ or $2$.
%%%%%%%
\item
$R$ is {\rm ($\NEG\LAND$E)}: 
{\scriptsize
$$
\infer[R']{\de}{
   \infer[(\NEG\LAND {\rm E})]{\ga}{
       \infer*[{\mathcal{D}}_1]{\NEG(\al\LAND\be)}{
        }
        &
       \infer*[{\mathcal{D}}_2]{\ga}{
          [\NEG\al]
        }
        &
       \infer*[{\mathcal{D}}_3]{\ga}{
          [\NEG\be]
        }
   }  
   &
   \infer*[{\mathcal{E}}_1]{\de_1}{
   }
   &
   \infer*[{\mathcal{E}}_2]{\de_2}{
   }
}
\quad\quad \gg \quad\quad 
\infer[(\NEG\LAND {\rm E})]{\de}{ 
   \infer*[{\mathcal{D}}_1]{\NEG(\al\LAND\be)}{
   }
   &
   \infer[\scriptstyle\!{R'}]{\de}{
      \infer*[{\mathcal{D}}_2]{\ga}{
          [\NEG\al]
       }
       &
       \infer*[{\mathcal{E}}_1]{\de_1}{
       }
       &
       \infer*[{\mathcal{E}}_2]{\de_2}{
       }
   }
   &
   \infer[\scriptstyle\!{R'}]{\de}{
      \infer*[{\mathcal{D}}_3]{\ga}{
          [\NEG\be]
       }
       &
       \infer*[{\mathcal{E}}_1]{\de_1}{
       }
       &
       \infer*[{\mathcal{E}}_2]{\de_2}{
       }
   }
}
$$}where $R'$ is an arbitrary inference rule, and both ${\mathcal{E}}_1$ and ${\mathcal{E}}_2$ are derivations of the minor premises of $R'$ if they exist. 
%%%%%%%%%
\item
$R$ is {\rm ($\NEG\LOR$I)} and $\ga$ is $\NEG(\al_1\LOR\al_2)$: 
{\footnotesize
$$
\infer[(\NEG\LOR {\rm E}i)]{\NEG\al_i}{
   \infer[(\NEG\LOR {\rm I})]{\NEG(\al_1\LOR\al_2)}{
       \infer*[{\mathcal{D}}_1]{\NEG\al_1}{
        }
        &
       \infer*[{\mathcal{D}}_2]{\NEG\al_2}{
        }
   }  
}
\quad\quad \gg \quad\quad 
   \infer*[{\mathcal{D}}_i]{\NEG\al_i}{
   }
\quad 
where~i~is~1~or~2. 
$$}
%where $i$ is $1$ or $2$.
%%%%%%%
\item
The set of derivations is closed under $\gg$. 
\end{enumerate}

%%%%%%%%%%%%%%%%%%%%%%%%%%%%%%%%%%%%%%%%%%%%
\item
The definition of the reduction relation $\gg$ at $\ga$ in {\rm nC3} is obtained from the conditions for the reduction relation $\gg$ at $\ga$ in {\rm nC} by adding the following condition. 
\begin{enumerate}
%%%%%%%
\item
\label{em-conditions}
$R$ is {\rm (EM)} and $\ga$ is $\ga_1\I\ga_2$, $\ga_1\LAND\ga_2$, $\ga_1\LOR\ga_2$, $\NEG\NEG\ga'$, $\NEG(\ga_1\I\ga_2)$, $\NEG(\ga_1\LAND\ga_2)$, or $\NEG(\ga_1\LOR\ga_2)$: 
{\scriptsize
$$
\infer[R']{\de}{
   \infer[({\rm EM})]{\ga}{
       \infer*[{\mathcal D}_1]{\ga}{
          [\NEG\al]
        }
        &
       \infer*[{\mathcal D}_2]{\ga}{
          [\al]
        }
   }  
   &
   \infer*[{\mathcal E}_1]{\de_1}{
   }
   &
   \infer*[{\mathcal E}_2]{\de_2}{
   }
}
\quad\quad \gg \quad\quad 
\infer[({\rm EM})]{\de}{ 
   \infer[R']{\de}{
      \infer*[{\mathcal D}_1]{\ga}{
          [\NEG\al]
       }
       &
       \infer*[{\mathcal E}_1]{\de_1}{
       }
       &
       \infer*[{\mathcal E}_2]{\de_2}{
       }
   }
   &
   \infer[R']{\de}{
      \infer*[{\mathcal D}_2]{\ga}{
          [\al]
       }
       &
       \infer*[{\mathcal E}_1]{\de_1}{
       }
       &
       \infer*[{\mathcal E}_2]{\de_2}{
       }
   }
}
$$}where $R'$ is 
%an arbitrary inference rule, 
{\rm ($\I$E)}, 
{\rm ($\LAND$E1)}, 
{\rm ($\LAND$E2)},
{\rm ($\LOR$E)},
{\rm ($\NEG\NEG$E)}
{\rm ($\NEG\I$E)},
{\rm ($\NEG\LAND$E)},
{\rm ($\NEG\LOR$E1)}, or 
{\rm ($\NEG\LOR$E2)},
and both ${\mathcal E}_1$ and ${\mathcal E}_2$ are derivations of the minor premises of $R'$ if they exist. 
\end{enumerate}

%%%%%%%%%%%%%%%%%%%%%%%%%%%%%%%%%%%%%%%%%%%%
\item
The definition of the reduction relation $\gg$ at $\ga$ in {\rm nMC} is obtained from the conditions for the reduction relation $\gg$ at $\ga$ in {\rm nC} by adding the following condition. 
\begin{enumerate}
%%%%%%
\item
$R$ is {\rm (GEM)} and $\ga$ is $\ga_1\I\ga_2$, $\ga_1\LAND\ga_2$, $\ga_1\LOR\ga_2$, $\NEG\NEG\ga'$, $\NEG(\ga_1\I\ga_2)$, $\NEG(\ga_1\LAND\ga_2)$, or $\NEG(\ga_1\LOR\ga_2)$: 
{\scriptsize
$$
\infer[R']{\de}{
   \infer[({\rm GEM})]{\ga}{
       \infer*[{\mathcal D}_1]{\ga}{
          [\al\I\be]
        }
        &
       \infer*[{\mathcal D}_2]{\ga}{
          [\al]
        }
   }  
   &
   \infer*[{\mathcal E}_1]{\de_1}{
   }
   &
   \infer*[{\mathcal E}_2]{\de_2}{
   }
}
\quad\quad \gg \quad\quad 
\infer[({\rm GEM})]{\de}{ 
   \infer[R']{\de}{
      \infer*[{\mathcal D}_1]{\ga}{
          [\al\I\be]
       }
       &
       \infer*[{\mathcal E}_1]{\de_1}{
       }
       &
       \infer*[{\mathcal E}_2]{\de_2}{
       }
   }
   &
   \infer[R']{\de}{
      \infer*[{\mathcal D}_2]{\ga}{
          [\al]
       }
       &
       \infer*[{\mathcal E}_1]{\de_1}{
       }
       &
       \infer*[{\mathcal E}_2]{\de_2}{
       }
   }
}
$$}where $R'$ is 
%an arbitrary inference rule, 
{\rm ($\I$E)}, 
{\rm ($\LAND$E1)}, 
{\rm ($\LAND$E2)},
{\rm ($\LOR$E)},
{\rm ($\NEG\NEG$E)}
{\rm ($\NEG\I$E)},
{\rm ($\NEG\LAND$E)},
{\rm ($\NEG\LOR$E1)}, or 
{\rm ($\NEG\LOR$E2)},
and both ${\mathcal E}_1$ and ${\mathcal E}_2$ are derivations of the minor premises of $R'$ if they exist. 
\end{enumerate}

%%%%%%%%%%%%%%%%%%%%%%%%%%%%%%%%%%%%%%%%%%%%
\item
The definition of the reduction relation $\gg$ at $\ga$ in {\rm nCN} is obtained from the conditions for the reduction relation $\gg$ at $\ga$ in {\rm nC3} by adding the other conditions of {\rm nMC}. Namely,  it is defined as all the conditions for both {\rm nC3} and {\rm nMC}. 
%%%%% 
\end{enumerate}
\end{df}

Prior to proving the normalization theorems for nC, nC3, nMC, and nCN, we need the following lemma.

\begin{lm}
\label{nc-systems-lemma}
Let $N_1$, $N_2$, $N_3$, and $N_4$ be {\rm nC}, {\rm nC3}, {\rm nMC}, and {\rm nCN}, respectively. 
Let $S_1$, $S_2$, $S_3$, and $S_4$ be {\rm sC}, {\rm sC3}, {\rm sMC$^*$}, and {\rm sCN$^*$}, respectively. 
For any $i \in \{1, 2, 3, 4\}$, the following hold.  
\begin{enumerate}
\item 
If $\mathcal{D}$ is a derivation in $N_i$ such that 
{\rm oa($\mathcal{D}$)} $=$ $\GA$ and 
{\rm end($\mathcal{D}$)} $=$ $\be$, 
then $S_i$ $\vdash$ \SEQ{\GA}{\be},

\item 
If $S_i$ $-$ {\rm (cut)} $\vdash$ \SEQ{\GA}{\be},  
then we can obtain a derivation $\mathcal{D}'$ in $N_i$ such that 
%\begin{enumerate}
%\item
(a) {\rm oa($\mathcal{D}'$)} $=$ $\GA$, 
%
%\item
(b) {\rm end($\mathcal{D}'$)} $=$ $\be$, and 
%
%\item 
(c) $\mathcal{D}'$ is normal.
%\end{enumerate} 
\end{enumerate}
\end{lm}
\PROOF~
%We only prove this lemma for nC3 and sC3 because the other systems can be handled similarly. 
\begin{enumerate}
%%%%%%%%
\item
We prove 1 by induction on the derivations $\mathcal{D}$ of $N_i$ such that oa($\mathcal{D}$) = $\GA$ and end($\mathcal{D}$) = $\be$. We distinguish the cases according to the last inference of $\mathcal{D}$. We show some cases. 
\begin{enumerate}
%%%%%%%%%%%
\item
Case (EM): 
$\mathcal{D}$ is of the form: 
{\footnotesize
$$
\infer[{\rm (EM)}]{\ga}{
     \infer*[{\mathcal{D}}_1]{\ga}{
          [\NEG\al] \GA_1
     }
     &
     \infer*[{\mathcal{D}}_2]{\ga}{
          [\al] \GA_2
     }
}
$$}where oa($\mathcal{D}$) = $\GA_1 \cup\GA_2$ %(multiset  union) 
and end($\mathcal{D}$) = $\ga$. 
By induction hypothesis, 
we have $S_i$ $\vdash$ \SEQ{\NEG\al, \GA_1}{\ga} and $S_i$ $\vdash$ \SEQ{\al, \GA_2}{\ga}.
Then, we obtain the required fact 
$S_i$ $\vdash$ \SEQ{\GA_1, \GA_2}{\ga}:
{\footnotesize
$$
\infer[(\mbox{\rm ex-middle})]{\SEQ{\GA_1, \GA_2}{\ga}}{
      \infer*[(\mbox{\rm we})]{\SEQ{\NEG\al, \GA_1, \GA_2}{\ga}}{
          \infer*[Ind.\, hyp.]{\SEQ{\NEG\al, \GA_1}{\ga}}{
          }
      }
      &
      \infer*[(\mbox{\rm we})]{\SEQ{\al, \GA_1, \GA_2}{\ga}}{
           \infer*[Ind.\, hyp.]{\SEQ{\al, \GA_2}{\ga}}{
           }
      }
}
$$
}where (we) is admissible in $S_i$ $-$ (cut) by Propositions \ref{we-prop} and \ref{we-prop-2}.
%Proposition \ref{we-prop}. 
%%%%%%%%%%%
%\item
%Case ($\NEG\NEG$E): 
%$\mathcal{D}$ is of the form: 
%{\footnotesize
%$$
%\infer[(\NEG\NEG {\rm E})]{\al}{
%     \infer*[{\mathcal{D}}_1]{\NEG\NEG\al}{
%          \GA
%     }
%}
%$$}where oa($\mathcal{D}$) = $\GA$ and end($\mathcal{D}$) = $\al$. 
%By induction hypothesis, 
%we have $S_i$ $\vdash$ \SEQ{\GA}{\NEG\NEG\al}. Then, we obtain the required fact 
%$S_i$ $\vdash$ \SEQ{\GA}{\al}:
%{\footnotesize
%$$
%\infer[({\rm cut}).]{\SEQ{\GA}{\al}}{
%       \infer*[Ind.\, hyp.]{\SEQ{\GA}{\NEG\NEG\al}}{
%       }
%       &
%       \infer[(\NEG\NEG {\rm left})]{\SEQ{\NEG\NEG\al}{\al}}{
%               \infer*[Prop. \ref{initial-sequent-prop}]{\SEQ{\al}{\al}}{
%               }
%       }
%}
%$$
%}
%%%%%%%%%%%%%%%%%
%\color{green}
\item
Case ($\NEG\I$I): 
We divide this case into two subcases. 
%%%%%%%%
\begin{enumerate}
\item
Subcase 1: 
$\mathcal{D}$ is of the form: 
{\footnotesize
$$
\infer[{\rm (\NEG\I I)}]{\NEG(\al\I\be)}{
     \infer*[{\mathcal{D}}']{\NEG\be}{
          \GA
     }
}
$$}where oa($\mathcal{D}$) = $\GA$ and end($\mathcal{D}$) = $\NEG(\al\I\be)$. 
By induction hypothesis, 
we have $S_i$ $\vdash$ \SEQ{\GA}{\NEG\be}. 
Then, we obtain that
$S_i$ $\vdash$ \SEQ{\GA}{\NEG(\al\I\be)}:
{\footnotesize
$$
\infer[(\NEG\I {\rm right})]{\SEQ{\GA}{\NEG(\al\I\be)}}{
     \infer[\mbox{\rm (we)}]{\SEQ{\al, \GA}{\NEG\be}}{
          \infer*[Ind. \, hyp.]{\SEQ{\GA}{\NEG\be}}{
          }
     }
}
$$
}where (we) is admissible in $S_i$ $-$ (cut) by Propositions \ref{we-prop} and \ref{we-prop-2}.
%Proposition \ref{we-prop}.
%%%%%%%%%%%
\item
Subcase 2: 
%Case ($\NEG\I$I): 
$\mathcal{D}$ is of the form: 
{\footnotesize
$$
\infer[(\NEG\I{\rm I})]{\NEG (\al\I\be)}{
     \infer*[{\mathcal{D}'}]{\NEG\be}{
          [\al]~\GA
     }
}
$$}where oa($\mathcal{D}$) = $\GA$ and end($\mathcal{D}$) = $\NEG (\al\I\be)$. 
By induction hypothesis, 
we have $S_i$ $\vdash$ \SEQ{\al, \GA}{\NEG\be}.
Then, we obtain the required fact 
$S_i$ $\vdash$ \SEQ{\GA}{\NEG (\al\I\be)}:
{\footnotesize
$$
\infer[(\NEG\I {\rm right}).]{\SEQ{\GA}{\NEG(\al\I\be)}}{
    \infer*[Ind. \, hyp.]{\SEQ{\al, \GA}{\NEG\be}}{
    }
}
$$
}
\end{enumerate}

%%%%%%%%%%%
\item
Case ($\NEG\I$E): 
$\mathcal{D}$ is of the form: 
{\footnotesize
$$
\infer[(\NEG\I{\rm E})]{\NEG\be}{
     \infer*[{\mathcal{D}}_1]{\NEG(\al\I\be)}{
          \GA_1
     }
     &
     \infer*[{\mathcal{D}}_2]{\al}{
          \GA_2
     }
}
$$}where oa($\mathcal{D}$) = $\GA_1 \cup\GA_2$ and end($\mathcal{D}$) = $\NEG\be$. 
By induction hypotheses, 
we have $S_i$ $\vdash$ \SEQ{\GA_1}{\NEG(\al\I\be)} and 
$S_i$ $\vdash$ \SEQ{\GA_2}{\al}.
Then, we obtain the required fact 
$S_i$ $\vdash$ \SEQ{\GA_1, \GA_2}{\NEG\be}:
{\footnotesize
$$
\infer[({\rm cut}).]{\SEQ{\GA_1, \GA_2}{\NEG\be}}{
   \infer*[Ind. \, hyp.]{\SEQ{\GA_2}{\al}}{
   }
   &
   \infer[({\rm cut})]{\SEQ{\al, \GA_1}{\NEG\be}}{
        \infer*[Ind. \, hyp.]{\SEQ{\GA_1}{\NEG(\al\I\be)}}{
        }
        &
        \infer[(\NEG\I {\rm left})]{\SEQ{\NEG(\al\I\be), \al}{\NEG\be}}{
              \infer*[Prop. \ref{initial-sequent-prop}]{\SEQ{\al}{\al}}{
              }
              &
              \infer*[Prop. \ref{initial-sequent-prop}]{\SEQ{\NEG\be}{\NEG\be}}{
              }
        }
   }
}
$$
}
%%%%%%%%%%%
\item
Case ($\NEG\LAND$E): 
$\mathcal{D}$ is of the form: 
{\footnotesize
$$
\infer[(\NEG\LAND{\rm E})]{\ga}{
     \infer*[{\mathcal{D}}_1]{\NEG(\al\LAND\be)}{
          \GA_1
     }
     &
     \infer*[{\mathcal{D}}_2]{\ga}{
          [\NEG\al] \GA_2
     }
     &
     \infer*[{\mathcal{D}}_3]{\ga}{
          [\NEG\be] \GA_3
     }
}
$$}where oa($\mathcal{D}$) = $\GA_1 \cup\GA_2 \cup \GA_3$ and end($\mathcal{D}$) = $\ga$. 
By induction hypotheses, 
we have 
$S_i$ $\vdash$ \SEQ{\GA_1}{\NEG(\al\LAND\be)},
$S_i$ $\vdash$ \SEQ{\NEG\al, \GA_2}{\ga}, and 
$S_i$ $\vdash$ \SEQ{\NEG\be, \GA_3}{\ga}. 
Then, we obtain the required fact 
$S_i$ $\vdash$ \SEQ{\GA_1, \GA_2, \GA_3}{\ga}:
{\footnotesize
$$
\infer[({\rm cut})]{\SEQ{\GA_1, \GA_2, \GA_3}{\ga}}{
    \infer*[Ind. \, hyp.]{\SEQ{\GA_1}{\NEG(\al\LAND\be)}}{
     }
    &
    \infer[(\NEG\LAND {\rm left})]{\SEQ{\NEG(\al\LAND\be), \GA_2, \GA_3}{\ga}}{
          \infer*[({\rm we})]{\SEQ{\NEG\al, \GA_2, \GA_3}{\ga}}{
               \infer*[Ind. \, hyp.]{\SEQ{\NEG\al, \GA_2}{\ga}}{
               }
          }
          &
           \infer*[({\rm we})]{\SEQ{\NEG\be, \GA_2, \GA_3}{\ga}}{
               \infer*[Ind. \, hyp.]{\SEQ{\NEG\be, \GA_3}{\ga}}{
               }
          }
    }
}
$$
}where (we) is admissible in $S_i$ $-$ (cut) by Propositions \ref{we-prop} and \ref{we-prop-2}.
%%%%%%%%%%%
%\item
%Case ($\NEG\LOR$I): 
%$\mathcal{D}$ is of the form: 
%{\footnotesize
%$$
%\infer[(\NEG\LOR{\rm I})]{\NEG (\al\LOR\be)}{
%     \infer*[{\mathcal{D}}_1]{\NEG\al}{
%          \GA_1
%     }
%     &
%     \infer*[{\mathcal{D}}_2]{\NEG\be}{
%          \GA_2
%     }
%}
%$$}where oa($\mathcal{D}$) = $\GA_1 \cup\GA_2$ and end($\mathcal{D}$) = $\NEG (\al\LOR\be)$. 
%By induction hypotheses, 
%we have $S_i$ $\vdash$ \SEQ{\GA_1}{\NEG\al} and 
%$S_i$ $\vdash$ \SEQ{\GA_2}{\NEG\be}.
%Then, we obtain the required fact 
%$S_i$ $\vdash$ \SEQ{\GA_1, \GA_2}{\NEG (\al\LOR\be)}:
%{\footnotesize
%$$
%\infer[(\NEG\LOR {\rm right}).]{\SEQ{\GA_1, \GA_2}{\NEG (\al\LOR\be)}}{
%      \infer*[({\rm we})]{\SEQ{\GA_1, \GA_2}{\NEG\al}}{
%          \infer*[Ind.\, hyp.]{\SEQ{\GA_1}{\NEG\al}}{
%          }
%      }
%      &
%      \infer*[({\rm we})]{\SEQ{\GA_1, \GA_2}{\NEG\be}}{
%           \infer*[Ind.\, hyp.]{\SEQ{\GA_2}{\NEG\be}}{
%           }
%      }
%}
%$$
%}
\end{enumerate}

%%%%%%%%%%%%%
\item
We prove 2 by induction on the derivations $\mathcal{D}$ of \SEQ{\GA}{\be} in $S_i$ $-$ (cut). We distinguish the cases according to the last inference of $\mathcal{D}$. We show some cases. 
\begin{enumerate}
%%%%%%%@deleted critical %%%%%%%%%%%
%\color{green}
%\item
%Case (we): 
%$\mathcal{D}$ is of the form:
%{\footnotesize
%$$
%\infer[\mbox{\rm (we-left)}]{\SEQ{\al, \GA}{\ga}}{
%  \infer*[\mathcal{E}]{\SEQ{\GA}{\ga}}{
%  }
%}
%$$}By induction hypothesis, we have a normal derivation ${\mathcal{E}}'$ in $N_i$ of the form: 
%{\footnotesize
%$$
%\infer*[{\mathcal{E}}']{\ga}{
%    \GA
%}
%$$}where oa(${\mathcal{E}}'$) = $\GA$ and end(${\mathcal{E}}'$) = $\ga$. 
%Then, we obtain a required normal derivation $\mathcal{D}'$ by: 
%{\footnotesize
%$$
%\infer[{\rm (\I E)}]{\ga}{
%    \infer[{\rm (Wk)}]{\al\I\ga}{
%         \infer*[{\mathcal{E}}']{\ga}{
%             \GA
%         }
%     }
%    &
%    \al
%}
%$$}where oa($\mathcal{D}'$) = $\{\al\}\cup\GA$ and end($\mathcal{D}'$) = $\ga$.
%We remark that this forms a normal derivation because (Wk) is not an introduction rule (i.e., $\al\I\ga$ is not a maximum formula). 
%\color{black}
%%%%%%%%%%%%%%%%%%%%%%%%%
%%%%%%%%%%
\item
Case (ex-middle): 
$\mathcal{D}$ is of the form: 
{\footnotesize
$$
\infer[(\mbox{\rm ex-middle})]{\SEQ{\GA}{\ga}}{
  \infer*[{\mathcal{D}}_1]{\SEQ{\NEG\al, \GA}{\ga}}{
  }
  &
  \infer*[{\mathcal{D}}_2]{\SEQ{\al, \GA}{\ga}}{
  }
}
$$}By induction hypotheses, we have normal derivations ${\mathcal{E}}_1$ and ${\mathcal{E}}_2$ in $N_i$ of the form: 
{\footnotesize
$$
\infer*[{\mathcal{E}}_1]{\ga}{
    \NEG\al, \GA
}
\quad\quad\quad
\infer*[{\mathcal{E}}_2]{\ga}{
    \al, \GA
}
$$}where 
oa(${\mathcal{E}}_1$) = $\{ \NEG\al \} \cup \GA$, 
oa(${\mathcal{E}}_2$) = $\{ \al \} \cup \GA$, 
end(${\mathcal{E}}_1$) = $\ga$, and end(${\mathcal{E}}_2$) = $\ga$.  
Then, we obtain a required normal derivation ${\mathcal{D}}'$ by:  
{\footnotesize
$$
\infer[({\rm EM})]{\ga}{
       \infer*[{\mathcal{E}}_1]{\ga}{
            [\NEG\al] \GA
       }
       &
       \infer*[{\mathcal{E}}_2]{\ga}{
            [\al] \GA
       }
}
$$}where oa(${\mathcal{D}}'$) = $\GA$ and end(${\mathcal{D}}'$) = $\ga$.

%%%%%%
\item
Case ($\NEG\NEG$left): 
$\mathcal{D}$ is of the form: 
{\footnotesize
$$
\infer[(\NEG\NEG {\rm left})]{\SEQ{\NEG\NEG\al, \GA}{\ga}}{
  \infer*[P']{\SEQ{\al, \GA}{\ga}}{
  }
}
$$}By induction hypothesis, we have a normal derivation $Q'$ in $N_i$ of the form: 
{\footnotesize
$$
\infer*[Q']{\ga}{
    \al, \GA
}
$$}where oa($Q'$) = $\{\al\} \cup \GA$ and end($Q'$) = $\ga$. 
Then, we obtain a required normal derivation $\mathcal{D}'$ by:  
{\footnotesize
$$
\infer*[Q']{\ga}{
    \infer[(\NEG\NEG {\rm E})]{\al}{
             \NEG\NEG\al
    }
    &
    \GA
}
$$}where oa($\mathcal{D}'$) = $\{ \NEG\NEG\al \} \cup \GA$ and end($\mathcal{D}'$) = $\ga$.

%%%%%%
\item
Case ($\NEG\I$left): 
$\mathcal{D}$ is of the form: 
{\footnotesize
$$
\infer[(\NEG\I {\rm left})]{\SEQ{\NEG(\al\I\be), \GA, \DE}{\ga}}{
  \infer*[{\mathcal{D}}_1]{\SEQ{\GA}{\al}}{
  }
  &
  \infer*[{\mathcal{D}}_2]{\SEQ{\NEG\be, \DE}{\ga}}{
  }
}
$$}By induction hypotheses, we have normal derivations ${\mathcal{E}}_1$ and ${\mathcal{E}}_2$ in $N_i$ of the form: 
{\footnotesize
$$
\infer*[{\mathcal{E}}_1]{\al}{
    \GA
}
\quad\quad\quad
\infer*[{\mathcal{E}}_2]{\ga}{
    \NEG\be, \DE
}
$$}where 
oa(${\mathcal{E}}_1$) = $\GA$, 
oa(${\mathcal{E}}_2$) = $\{\be\}\cup\DE$, 
end(${\mathcal{E}}_1$) = $\al$, and 
end(${\mathcal{E}}_2$) = $\ga$.  
Then, we obtain a required normal derivation $\mathcal{D}'$ by:  
{\footnotesize
$$
\infer*[{\mathcal{E}}_2]{\ga}{
       \infer[(\NEG\I {\rm E})]{\NEG\be}{
              \NEG(\al\I\be)
              &
              \infer*[{\mathcal{E}}_1]{\al}{
                  \GA
              }
       }
       &
       \DE
}
$$}where oa($\mathcal{D}'$) = $\{\NEG(\al\I\be)\}\cup \GA \cup \DE$ and end($\mathcal{D}'$) = $\ga$.

%%%%%%
\item
Case ($\NEG\LAND$left): 
$\mathcal{D}$ is of the form: 
{\footnotesize
$$
\infer[(\NEG\LAND {\rm left})]{\SEQ{\NEG(\al\LAND\be), \GA}{\ga}}{
  \infer*[{\mathcal{D}}_1]{\SEQ{\NEG\al, \GA}{\ga}}{
  }
  &
  \infer*[{\mathcal{D}}_2]{\SEQ{\NEG\be, \GA}{\ga}}{
  }
}
$$}By induction hypotheses, we have normal derivations ${\mathcal{E}}_1$ and ${\mathcal{E}}_2$ in $N_i$ of the form: 
{\footnotesize
$$
\infer*[{\mathcal{E}}_1]{\ga}{
    \NEG\al, \GA
}
\quad\quad\quad
\infer*[{\mathcal{E}}_2]{\ga}{
    \NEG\be, \GA
}
$$}where 
oa(${\mathcal{E}}_1$) = $\{\NEG\al\}\cup\GA$,
oa(${\mathcal{E}}_2$) = $\{\NEG\be\}\cup\GA$, and  
end(${\mathcal{E}}_1$) = end(${\mathcal{E}}_2$) = $\ga$.  
Then, we obtain a required normal derivation $\mathcal{D}'$ by:  
{\footnotesize
$$
\infer[(\NEG\LAND{\rm E})]{\ga}{
      \NEG(\al\LAND\be)
      &
      \infer*[{\mathcal{E}}_1]{\ga}{
            [\NEG\al] \GA
      }
      &
      \infer*[{\mathcal{E}}_2]{\ga}{
            [\NEG\be] \GA
      }
}
$$}where oa($\mathcal{D}'$) = $\{\NEG(\al\LAND\be)\}\cup\GA$ and end($\mathcal{D}'$) = $\ga$.

%%%%%%
\item
Case ($\NEG\LOR$right): 
$\mathcal{D}$ is of the form: 
{\footnotesize
$$
\infer[(\NEG\LOR {\rm right})]{\SEQ{\GA}{\NEG (\al\LOR\be)}}{
  \infer*[{\mathcal{D}}_1]{\SEQ{\GA}{\NEG\al}}{
  }
  &
  \infer*[{\mathcal{D}}_2]{\SEQ{\GA}{\NEG\be}}{
  }
}
$$}By induction hypotheses, we have normal derivations ${\mathcal{E}}_1$ and ${\mathcal{E}}_2$ in $N_i$ of the form: 
{\footnotesize
$$
\infer*[{\mathcal{E}}_1]{\NEG\al}{
    \GA
}
\quad\quad\quad
\infer*[{\mathcal{E}}_2]{\NEG\be}{
    \GA
}
$$}where oa(${\mathcal{E}}_1$) = oa(${\mathcal{E}}_2$) = $\GA$, end(${\mathcal{E}}_1$) = $\NEG\al$, and end(${\mathcal{E}}_2$) = $\NEG\be$.  
Then, we obtain a required normal derivation $\mathcal{D}'$ by:  
{\footnotesize
$$
\infer[(\NEG\LOR{\rm I})]{\NEG (\al\LOR\be)}{
       \infer*[{\mathcal{E}}_1]{\NEG\al}{
            \GA
       }
       &
       \infer*[{\mathcal{E}}_2]{\NEG\be}{
            \GA
       }
}
$$}where oa($\mathcal{D}'$) = $\GA$ and end($\mathcal{D}'$) = $\NEG (\al\LOR\be)$. 
\end{enumerate}
\end{enumerate}
\QED
%\\

%We then obtain the following theorems.

\begin{thm}[Equivalence between nC-family and sC-family]
Let $N_1$, $N_2$, $N_3$, and $N_4$ be {\rm nC}, {\rm nC3}, {\rm nMC}, and {\rm nCN}, respectively.
Let $S_1$, $S_2$, $S_3$, and $S_4$ be {\rm sC}, {\rm sC3}, {\rm sMC$^*$}, and {\rm sCN$^*$}, respectively. 
For any formula $\al$ and any $i \in \{1, 2, 3, 4\}$, $S_i$ $\vdash$ \SEQ{}{\al} iff  $\al$ is provable in $N_i$.
\end{thm}
\PROOF
Taking $\emptyset$ as $\GA$ in Lemma \ref{nc-systems-lemma}, we obtain the claim. 
\QED

\begin{thm}[Normalization for nC, nC3, nMC, and nCN]
%nC-family]
\label{normal-nj-star}
Let $N$ be {\rm nC}, {\rm nC3}, {\rm nMC}, or {\rm nCN}.   
All derivations in $N$ are normalizable. 
More precisely, if a derivation $\mathcal{D}$ in $N$ is given, 
then we can obtain a normal derivation $\mathcal{D}'$ in $N$ such that 
{\rm oa($\mathcal{D}'$)} $=$ {\rm oa($\mathcal{D}$)} and 
{\rm end($\mathcal{D}'$)} $=$ {\rm end($\mathcal{D}$)}. 
\end{thm}
\PROOF
Let $N_1$, $N_2$, $N_3$, and $N_4$ be {\rm nC}, {\rm nC3}, {\rm nMC}, and {\rm nCN}, respectively.
Let $S_1$, $S_2$, $S_3$, and $S_4$ be {\rm sC}, {\rm sC3}, {\rm sMC$^*$}, and {\rm sCN$^*$}, respectively.
Let $i$ be $1$, $2$, $3$, or $4$. 
Suppose that a derivation $\mathcal{D}$ in $N_i$ is given, and suppose that oa($\mathcal{D}$) = $\GA$ and end($\mathcal{D}$) = $\be$. Then, by Lemma \ref{nc-systems-lemma} (1), we obtain $L_i$ $\vdash$ \SEQ{\GA}{\be}. By the cut-elimination theorem for $S_i$ (i.e., Theorems \ref{cut-eli-sC}, \ref{cut-eli-C-systems}, and \ref{cut-eli-sMC-star}), we obtain $S_i$ $-$ (cut) $\vdash$ \SEQ{\GA}{\be}. Then, by Lemma \ref{nc-systems-lemma} (2), we can obtain a normal derivation $\mathcal{D}'$ in $N_i$ such that oa($\mathcal{D}'$) = oa($\mathcal{D}$) and end($\mathcal{D}'$) = end($\mathcal{D}$). 
\QED
%\\

\

\noindent {\bf Acknowledgments.} 
%%%%%%% 2024/7/14 %%%%%%%%%
I would like to thank Heinrich Wansing and the members of the contradictory logics seminar at Ruhr University Bochum for their valuable comments and discussions, especially at the 1st Workshop on Contradictory Logics, held in Bochum from December 6 to 8, 2023. I would also like to thank the anonymous referees of NCL'24 for their valuable comments. This research was supported by JSPS KAKENHI Grant Number 23K10990.
%%%%%%%%%%%%%%%%%%%%%%%%%%%%

%%%%%%%%%%%%%%%%%%%%%%%%%%%%%%%%%%%%%%
\nocite{*}
\bibliographystyle{eptcs}
\bibliography{generic}

\end{document}